\documentstyle[11pt,rotate,epsfig]{article}

\def\text{~}

\def\hc2{(\hbar c)^2}

\def\r2{\langle r^2 \rangle}
\def\Q2{$Q^2$}

\def\gev2c2{GeV$^2$/$c^2$}
\def\fm2{\text{fm}^2}
\def\a2         {{\mbox{$a_{2}                                        \
$}}}

\def\a2pigamma  {{\mbox{$a_2^-\rightarrow \pi^-\gamma                \
$}}}

\begin{document}

 \pagestyle{empty}
 \begin{flushright}
 TAUP-2671-2001\\
 12 June 2001
 \end{flushright}
 
 \begin{center}
 
  \vspace*{3.5cm}
 {\LARGE {\bf Color Transparency at COMPASS}\footnote{
 Expanded version of the talk presented
 at the Workshop on "Nucleon Structure and Meson Spectroscopy"
 physics to be studied at the COMPASS experiment at CERN, 
 Dubna, Russia, 10--11 October 2000.
 The conference www site is http://wwwcompass.cern.ch/compass/workshops/dubna2000/index.html.}}\\
  
  \vspace*{3mm}
  
  {\Large --- Feasibility Study ---\\}
 
 \vspace*{1.8cm}

 \mbox{\Large Andrzej Sandacz$^{2} \!,$
 \,Oleg A. Grajek$^{2} \!$,
 \,Murray Moinester$^{3} \!$,
 \,Eli Piasetzky$^{3}$}

 \vspace*{8mm}

 $^{2}$ So\l tan Institute for Nuclear Studies, ul. Ho\.{z}a 69, PL 00-681 Warsaw,
 Poland,
 
 \vspace*{2mm}
 
 $^{3}$ School of Physics and Astronomy, R. and B. Sackler Faculty of Exact
 Sciences,\\ Tel Aviv University, 69978 Ramat Aviv, Israel

 \vspace*{4mm}

{\it E-mails: sandacz@fuw.edu.pl, oleg@fuw.edu.pl, murraym@tauphy.tau.ac.il, eip@tauphy.tau.ac.il}
 \end{center}
 
 \vspace*{2cm}
 
 \begin{abstract}
 
 \vspace*{0.4cm}
 
 We examine the potential of the COMPASS experiment at CERN to study color transparency
via exclusive vector meson production in hard muon-nucleus scattering.
It is demonstrated that COMPASS has high sensitivity to test this
important prediction of perturbative QCD.
 
 \vspace*{4.0cm}
 
 \end{abstract}

 \pagebreak
 
 \pagestyle{plain}
 \setcounter{page}{2}

 \tableofcontents
 \pagebreak

 \section{Introduction}
 \label{lab_sec_1}
 
\hspace*{8mm}{\it Color transparency\/} (CT) 
is a phenomenon
of perturbative QCD (pQCD),
whose characteristic feature is that small color-singlet objects
interact with hadrons with small cross sections \cite{low,bb,bfgms,fms}. 
Cross section for the interaction of such small object, or small size
configuration (SSC), with a hadron target has been calculated in QCD
using the factorization theorem \cite{bbfs,fms,fks,frs}.
These QCD calculations confirmed the hypothesis
of F. Low \cite{low} on smallness of the cross section for the
interaction of SSC with a hadron, if the gluon density in the
hadron is not very high (moderately small $x$).
They also predict a related phenomenon of {\it color opacity\/} 
when the gluon density becomes very large
(at small $x$) and SSC interacts with the hadron with large cross
section \cite {bb,fms}.

The prerequisite for observing CT is to select a sample containing SSC's 
via a hard process (i.e. with large $Q^{2}\!$, \,high $p^{}_{t}\/$,
or large produced mass).      
To suppress non-perturbative (not SSC) background different
additional restrictions should be imposed depending on the process.
For instance for hard exclusive $\rho ^0$ leptoproduction,
in addition to large $Q^2$, selection of the longitudinally
polarized mesons is required.
To 'measure' the SSC-nucleon cross sections one should 
study absorption of the SSC propagating through nuclear 
matter. In order to clearly observe CT it is necessary that the SSC 
lives long enough to traverse distances larger
than the size
of the target nucleus.
Another requirement is that SSC stays small
while propagating through the nucleus.
These requirements, which are quantified in terms of the coherence
length and the formation length, are discussed later.

Various processes have been proposed to study CT phenomenon.
\begin{itemize}
\item[(a)]
Coherent vector meson ($J \! / \! \psi\/$, $\rho $, $\phi $) 
production at small $t$.
For near-forward ($t \approx 0$) coherent production and complete CT, one expects
that the cross section for the nuclear target is related to that 
for the nucleon target by
\begin{equation}
\frac{{\rm d}\sigma}{{\rm d}t}(\gamma ^* A \rightarrow V A) \, |^{}_{t \approx 0} =
A^2 \frac{{\rm d}\sigma}{{\rm d}t}(\gamma ^* N \rightarrow V N) \, |^{}_{t \approx 0} \: .
\end{equation}
Gluon shadowing/antishadowing
in nuclei is neglected in this formula, but
for moderately small $x (> 0.01)$ this effect is expected to be small.
Usually in experiments the $t$-integrated coherent cross sections are 
measured, for which CT predicts an approximate $A^{4/3}$ dependence.
Using more realistic nuclear form factors one predicts $A^{1.40}$
\cite {Sokolov}.

Otherwise, if the color-singlet objects, while propagating in nuclear matter,
interact with nucleons with large cross sections, 
the expected $A$-dependence is weaker; 
$A^{2/3}$ for the cross sections comparable or larger than the pion-nucleon cross section.

\item[(b)]
Incoherent (quasi-elastic) vector meson ($J \! / \! \psi\/$, $\rho $, $\phi $) 
production on nuclei.
For complete CT and neglecting effects of the gluon shadowing/antishadowing
in the nuclei one expects
\begin{equation}
\frac{{\rm d}\sigma}{{\rm d}t}(\gamma ^* A \rightarrow V N (A-1)) =
A \frac{{\rm d}\sigma}{{\rm d}t}(\gamma ^* N \rightarrow V N) \: .
\end{equation}
Here, for the nuclear target
the meson $V\/$ is produced on a single nucleon of the nucleus and
$(A-1)$ denotes the system of spectator nucleons.

\item[(c)]
Coherent or incoherent production of excited vector meson states
$\psi '$ or $\rho '$.
CT predicts the same $A$ dependence of $J \! / \! \psi\/$ and $\psi '$,
or $\rho $ and $\rho '$. This contrasts with the naive expectations
whereby one may expect larger absorption for excited mesons
since they are larger. 

\item[(d)]
Coherent diffractive dissociation of hadrons or photons
into high $p_t$ di-jets. Such process probes the
small transverse-size component of the projectile wave-function as well as
CT effects. For CT the $t$-integrated 
cross section of coherent diffractive production of
high $p_t$ di-jets has the same $A$-dependence
as for the processes (a). 
Using more realistic wave functions one predicts
for asymptotically high energies $A^{\alpha }$ dependence,
where $\alpha $ is in the range 1.45 -- 1.60,
depending on $p_t$ \cite{fms}.

\item[(e)]
Coherent vector meson production on light nuclei (deuteron, helium)
in the large $t$ range, where the effects of double (multiple) scattering
are important. CT will suppress the double (multiple)
scattering contribution to
the differential cross section ${\rm d}\sigma /{\rm d}t$ \cite{fms,fgkss}.

\item[(f)]
$A$-dependence of the fraction of large rapidity gap (diffractive)
events. CT predicts an $A$ independent fraction; otherwise it
will grow with $A$ \cite{rapgap}.

\item[(g)]
Large $Q^2$ quasielastic $(e,e'p)$ scattering on nuclei,
$e A \rightarrow e p (A-1)$. In the CT limit the cross section
will be proportional to $A$, as for process (b) \cite{almu}.

\item[(h)]
Large $t$ quasielastic $(p,2p)$ scattering on nuclei,
$p A \rightarrow p p (A-1)$. Also for this process, for complete
CT the expected $A$ dependence is like for process (b) \cite{almu}.

\end{itemize}

In searches for CT 
a commonly used quantity, measured in experiments,
is the {\it nuclear transparency\/}
\begin{equation}
T = \frac{\sigma^{}_{\! \! A}}{A \, \sigma^{}_{\! N}} \: ,
\end{equation} 
which is the ratio of the cross section per nucleon for a selected process
on a nucleus $A$ to the corresponding cross section on a free nucleon.
For the incoherent processes CT predicts $T \simeq 1$ independently of $A$.
For the large absorption in the nuclear matter $T$ will be smaller 
than unity
and will decrease with $A$.
Although the nuclear transparency could be defined also for
coherent processes, usually cross sections for coherent production
on different nuclei are compared directly. As mentioned above,
for $t$-integrated coherent cross sections CT predicts 
$\sigma^{}_{\! \! A} \propto A^{4/3} \!$, \,whereas for a larger absorption the 
$A$-dependence is weaker.

\begin{figure}[b]
\begin{center}
\mbox{\epsfig{file=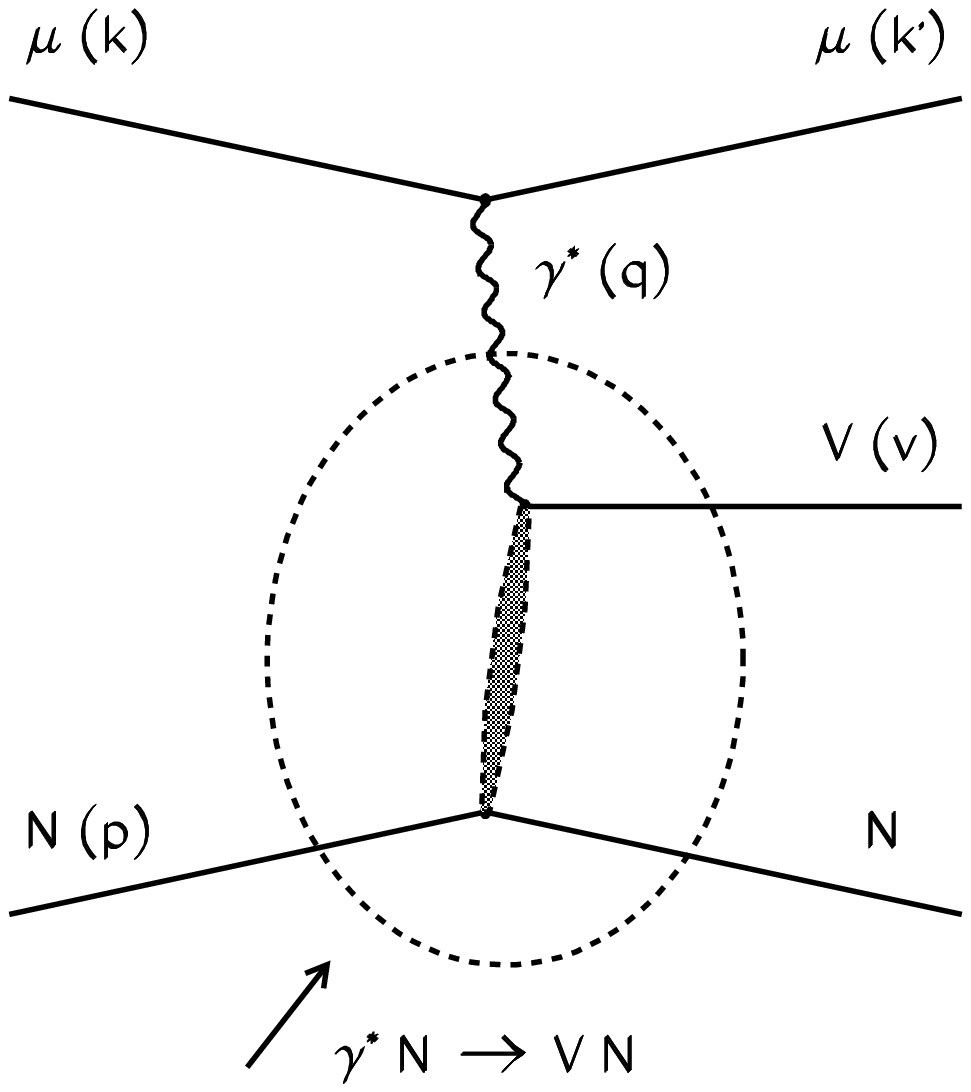,height=5.0in}}
\end{center}
\caption{Diagram of exclusive meson muoproduction
$\mu \: N \rightarrow \mu  \: N \: V $, where meson $V$
is the only produced particle. Corresponding four-momenta
are indicated in brackets (cf. Table 1)
. \label{diag}}
\end{figure}

In the following discussion we concentrate on 
exclusive vector meson production (VMP), the processes (a)\,--\,(c),
which could be studied at the COMPASS experiment
\cite{compass,bradamante}.
Exclusive meson production on the nucleon, free or bound in the nucleus,
can be viewed as proceeding according to the diagram shown in
Fig. \ref{diag}.
Kinematic variables used in this paper are listed in Table \ref{kinvar}.


\begin{table}[t]
\caption{Kinematic variables used in the text.
\label{kinvar}}
\vspace{0.2cm}
\begin{center}
\footnotesize
\begin{tabular}{|l|l|}
\hline
\hline
$k$         &   four-momentum of the incident muon,                      \\
$k'$        &   four-momentum of the scattered muon,                     \\
$p$         &   four-momentum of the target nucleon,                     \\
$v$         &   four-momentum of the vector meson $V$,                   \\
$q = k - k'$  &  four-momentum of the virtual photon,                   \\
--$Q^2 = q^2$ &  invariant mass squared of the virtual photon $\gamma^{\ast} \:$,           \\
$\nu  =(p\cdot q)/M^{}_{\! p}$ & energy of the virtual photon in the laboratory system,
\\
    & $M^{}_{\! p}$ is the proton mass,\\
$x = Q^2/(2 M^{}_{\! p} \nu )$ & Bjorken scaling variable,\\
$y = (p\cdot q)/(p\cdot k)$    & fraction of the lepton energy lost in the
laboratory system,\\
$W^2 = (p+q)^2$  & total energy squared in the $\gamma^{\ast} \! - \! N$ system,    \\
$t = (q-v)^2$ & four-momentum transfer to the target,\\
$p_t^2$      & transverse-momentum squared of the vector meson \\
   &  with respect to the virtual photon direction,\\
$m^{}_{\! V} = (v^2)^{\frac{1}{2}}$ & invariant mass of the vector meson $V$,\\
$M^{2}_{\! X} = (p+q-v)^2$ & missing-mass squared of the undetected
recoiled system,\\
$I = (M^{2}_{\! X} - M^{2}_{\! p})/W^2$ & inelasticity, \\
$z = (p\cdot v)/(p\cdot q)$ & fraction of the virtual photon energy in the laboratory sy- \\
   & stem taken by meson $V$, \\
   & $z \approx 1-I $\\
\hline
\hline
\end{tabular}
\end{center}
\end{table}

One of predictions of pQCD is that at high $Q^2$
the longitudinally polarized virtual photons $\gamma^{\ast}_{L}\/$
fluctuate into hadronic components, e.g. $q\bar{q}$ pairs,
whose transverse size 
$b = \linebreak \: \mid \! \bar{r}_{\! \perp q} - \bar{r}_{\! \perp \bar{q}} \! \mid $ 
decreases with $Q^2 \!$, \,$b \propto (Q^{2})^{-1/2} \!$.
\,At large $Q^2$ the values of $b$ are 
significantly smaller than the size of the nucleon.
For instance, at $Q^2 = 10 \: \rm{GeV}^2$ and at $x = 0.01$
$b_{u\bar{u}} \simeq b_{d\bar{d}} \simeq 0.3 \;$fm
\cite{fks}, to be compared
with the size of $\rho $ meson $b_{\! \rho } \simeq 1.4 \;$fm.
For transversely polarized photons, in addition to the small size
fluctuations, non-perturbative large size components
may be expected, even at reasonable large $Q^2$ $(\simeq 10 \:\rm{GeV}^2)$.

Another pQCD prediction is that the heavy quarks fluctuations, e.g.
$c\bar{c}$, of the virtual photon have small transverse size already  
for quasi-real production;  
$b_{c\bar{c}} = 0.22 \;$fm at $Q^2 \simeq 0$. 
 
The total cross section for the interaction of a small size
$q\bar{q}$ pair with the nucleon is given in pQCD by the formula \cite{bbfs}
\begin{equation}
 \sigma _{q\bar{q}, \: N} = \frac{\pi^2}{3} \: b^2 \: \alpha_s(Q^2) \; x \; g(x, \: Q^2) \;,
\end{equation}
where $g(x, \: Q^2)$ is the gluon distribution function in the nucleon.
The cross section for the interaction of $q\bar{q}$ pair of $b = 0.3 \;$fm with the nucleon 
is about 3$\;$mb at $x = 10^{-2}\,$ \cite{fks}.
At very small $x$ the gluon distribution $g\/$ in the nucleon increases,
which leads to an increase of the cross section $\sigma _{q\bar{q}, \: N}$.
For instance, for $b = 0.3 \;$fm $\sigma _{q\bar{q}, \: N}$ is
about 18$\;$mb at $x = 10^{-5} \!$.
\,Therefore, even for the small $q\bar{q}$ objects the cross section
becomes large at sufficiently small $x$. This phenomenon is mentioned earlier
{\it color opacity\/} and it could be studied at small $x$ at 
future electron-nucleus colliders.

In order to study CT for the exclusive production of light quark mesons
(like $\rho $ or $\rho'$)
one should select {\it large\/} $Q^2 \!$,
\,{\it moderately small\/} $x$ and {\it longitudinally\/} polarized mesons.
For similar studies for the production of $J \! / \! \psi\/$ or $\psi '$
one should just select {\it moderately small\/} $x$.
 
In studies of CT an important role is played by the {\it coherence length\/}
and the {\it formation length\/}. The coherence length $l_c$ (sometimes referred to
as propagation or interaction length) is defined
as the distance traversed by the $q\bar{q}$ fluctuation of the virtual photon
in the target nucleon/nucleus system and is given by
\begin{equation} 
l_c = \frac{2 \nu}{Q^2+M^2} = \frac{\beta }{M^{}_{\! N} \: x} \; .
\end{equation}
Here $\beta = Q^2 / (Q^2+M^2)$, $M$ is the invariant mass of the
$q\bar{q}$ fluctuation, and $M^{}_{N}\/$ is the mass of the nucleon.
For $l^{}_{c}\/$ values smaller than the size of the nucleus 
the life-time of the $q\bar{q}$ fluctuation is short and
a chance for
the hadronic fluctuation to interact in  
nuclear matter decreases. In particular this significantly inhibits
coherent production. For incoherent production the effects due to the
small $l^{}_{c}\/$ values mimic  the $Q^2$ dependence of the
nuclear transparency predicted by CT \cite{RPW}.

The formation length is the distance in the target system
needed for the $q\bar{q}$ fluctuation,
which scattered on a nucleon,
to develop into a hadron $h$. It is equal to
\begin{equation}
l^{}_{\! f} = \frac{\nu }{m^{}_{h} \: \Delta m} \; ,
\end{equation}
where $m^{}_{h}\/$ is the mass of the hadron $h$ and $\Delta m$ is the mass difference
between the hadron $h$ and its lowest orbital excitation.
For small
$l^{}_{\! f}\/$ values the $q\bar{q}$ fluctuation evolves quickly into
a full-size final hadron and
the absorption of the final hadron $h$ in the nucleus plays a role.

Experimentally the effects of the coherence length and the formation
length were analysed in detail by the HERMES experiment 
for incoherent exclusive production of $\rho^{0}\/$ on nuclei \cite{HERMES}
and were shown to play an important role at the small $l^{}_{c}\/$ values.

Therefore, for a clean demonstration of CT effects the optimal conditions are when the
values of $l^{}_{c}\/$ and $l^{}_{\! f}\/$ exceed the
size of the target nucleus. If the above is not feasible, 
the variation of nuclear transparency
$T(A,Q^2)$ with $Q^2$ should be studied for different, fixed
values of coherence length. This way, a change of $T(A)$ between
low and high $Q^2$ values could be associated with the onset of CT,
and not with varying $l^{}_{c}\/$.

The $t$-dependence of the cross section for exclusive meson production
on the nucleon is approximately exponential,
$e^{b(t-t_{min})}$, where
$t_{min} \approx -M_p^2x^2(1+ M^2/Q^2)^2$ and $M$ is the mass of the produced
meson. For soft QCD
processes the slope parameter $b$ should be independent of $Q^2$, 
depending only on the energy. This results from the Regge model \cite{collins}.
For hard VMP, if SSC's are important,
one predicts \cite{Fra98} that
the slope $b$ will significantly decrease with increasing $Q^2$
approaching a universal value related to the nucleon radius.
This is so
because the transverse interquark distances in the SSC decrease with
increasing $Q^2$. This prediction 
agrees with high energy
data on hard diffractive vector meson production on the proton
\cite {h1,zeus,Fra98b}. 

For VMP on a nucleus $A$ the $t$-dependence of the cross section
is approximately reproduced by a sum of two exponential functions. 
The peak at the lowest $t$ values,
with the slope 
proportional to the nucleus squared radius $< \! R^{2}_{\! A} \! >$,
is mostly due to the coherent production, whereas at somewhat larger 
$t$ the incoherent production on quasi-free nucleons dominates
and the slope is equal to that for the production on free nucleons.
Applying cuts on $t$ allows to select samples of events,
which are strongly dominated by either coherent
or incoherent production. It was successfully demonstrated in
Refs \cite{Sokolov,NMCJpsi,Wei97}.

In order to observe maximal
coherent exclusive meson production for which the whole
nucleus contributes,
it is necessary to satisfy the
condition 
\begin{equation}
|t_{min}< \! R^{2}_{\! A} \! >\!/3\, | << 1 \:.  
\end{equation}
In the COMPASS experiment for low $x$ values in the range 0.006 -- 0.02
 and for $Q^2 \approx 2-10 ~\:\rm{GeV}^2$, one has
$|t_{min}< \! R^{2}_{\! A} \! >\!/3\, |
 \approx ~10x^2 A^{2/3}$ and the condition (7) is satisfied 
even for lead $(A=207)$.
 
Experimental searches for CT started more than a decade ago and 
encompass various processes:
large $t$ quasielastic $(p, \: 2p)$ scattering \cite{AGS},
large $Q^2$ quasielastic $(e, \: e'p)$ scattering \cite{SLAC},
$J \! / \! \psi\/$ photoproduction \cite{Sokolov} and $J \! / \! \psi\/$
muoproduction \cite{NMCJpsi},
exclusive $\rho^{0}$ leptoproduction \cite{E665,nmc,HERMES}
and coherent diffractive dissociation of the pion into two high-$p^{}_{t}\/$ jets
\cite{Wei97}.

The pioneering studies of CT in large $t$ quasi-elastic proton
scattering \cite{AGS} found a rise of the nuclear transparency as the beam
energy increased from 6 to 9 GeV and a decrease at higher energies.
The large $Q^2$ quasi-elastic electron scattering studies \cite{SLAC}
did not show $Q^2$ dependence. The explanation of these results in terms
of CT is still debatable.

Strong recent evidence for CT comes from
Fermilab E791 experiment on the \mbox{$A$-de-} pendence
of coherent diffractive dissociation of pions into two high-$p^{}_{t}\/$ jets
\cite{Wei97}. Also the E691 results on $A$-dependence of coherent
$J \! / \! \psi\/$ photoproduction \cite{Sokolov} and the NMC measurements
of $A$-dependence of coherent and quasielastic $J \! / \! \psi\/$ muoproduction
are consistent with CT. Measurements of the nuclear transparency
for incoherent exclusive $\rho^{0}$ production by Fermilab experiment E665
give a hint for CT. However, due to the low statistics of that data at
high $Q^2 \!$,
\,it was not possible to disentangle effects of decreasing
$l^{}_{c}\/$ at high $Q^2$ and to demonstrate CT unambiguously.

\section{Experimental method}

\label{lab_sec_2}

\hspace*{8mm}We propose to study CT via {\it exclusive vector meson production\/}
$\mu A \rightarrow \mu \, V A$ (coherent) and $\mu A \rightarrow \mu \, V N (A-1)$
(incoherent) on various nuclei $A$ and optionally also on a 
proton or deuteron target.
As a primary objective we propose to study the 
production of the following mesons $V$: $\rho^{0} \!$, \,$J \! / \! \psi\/$,
$\phi $,
$\psi' (\psi (2S))$ and $\rho' (\rho (1450), \rho (1700))$.
Also investigations of the production of other
mesons will be possible. 
The preferable decay modes are those into the charged particles: 
$\rho ^0 \rightarrow \pi^{+} \pi^{-} \!$,
\,$J \! / \! \psi \rightarrow e^{+} e^{-} / \mu^{+} \mu^{-} \!$,
\,$\phi \rightarrow K^+ K^- \!$,
\,$\psi ' \rightarrow J \! / \! \psi \,  \pi^{+} \pi^{-} \!$,
\,$\rho' \rightarrow \pi^{+} \pi^{+} \pi^{-} \pi^{-} \!$.
\,Two or more nuclear targets will be used. An additional
proton (deuteron) target would be beneficial.

Our proposal is to complement the initial setup
of the COMPASS \cite{compass,bradamante}
for the muon run with the polarized target, 
by adding two thin nuclear
targets of lead and carbon of 17.6 g/cm$^2$ each.
The carbon target will be a cylinder 8 cm long and of 3 cm of diameter.
The lead target will consist of 4 discs of 3 cm of diameter, distributed
over length of 8 cm. Only one nuclear target will be exposed to the 
muon beam at a time, with frequent exchanges (every few hours)
of different targets. The nuclear targets will be located downstream 
of the polarized target, at the end of the solenoid magnet and before
the first tracking detector (first micromega chamber).

The high-intensity high-energy incident muon-beam
will impinge on the polarized target and a downstream thin nuclear target.
The momenta of the scattered muon and of the produced
charged particles will be reconstructed in the
two magnetic spectrometers, using the magnets SM1 and SM2,
instrumented with micromega chambers, drift chambers, GEM detectors,
straw chambers, multiwire proportional chambers and scintillating
fibers.
Adding a recoil detector which will surround the nuclear target
and register slow
particles emitted from it may not be possible for the present setup
due to the limited space.
However, it would be advantageous and possible for dedicated runs
taken with a modified setup.
In the following we assume that the recoil particle(s) remain(s)
undetected, and in order to
select exclusive events one has to rely only on the kinematics of
the scattered muon and the produced meson. 

The trigger will use the information
from hodoscopes registering the scattered muons, and from calorimeters
registering deposits of energy of the particles in the final state.
For VMP reactions triggering only on the scattered muon
is in principle possible. However, for better efficiency of the trigger,
especially at small $Q^2 \!$,
\,it will be useful for certain processes 
to require in addition  a minimum energy deposit in the calorimeters.

\begin{figure}[t]

\begin{center}
\mbox{\epsfig{file=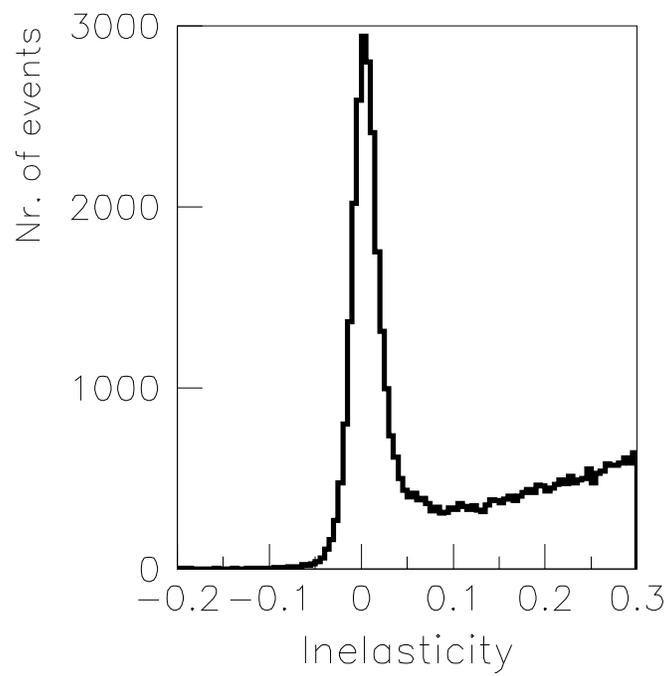,height=3.5in}} 
\end{center}

\caption{Inelasticity distribution.
The SMC preliminary results [29] for
$\mu N \rightarrow \mu \rho ^0 N$
. \label{ange}}
\end{figure}

The off-line selection of exclusive events for the production of different
mesons will be similar to that described for $\rho^{0}$ production 
in Ref. \cite{nmc,memo2000,pmmpsv}.
In particular the discrimination of non-exclusive events will be done
by applying cuts on the inelasticity $I\/$ (for the definition see
Table \ref{kinvar}). In Fig. \ref{ange} the inelasticity distribution is shown
for the SMC $\rho ^0$ sample ~\cite{smcrho}
for the events with the invariant mass in the
central part of the $\rho ^0$ peak.
For the inelasticity
distribution the peak at $I=0$ is the signal of exclusive $\rho ^0$
production. Non-exclusive events, where in addition to detected
fast hadrons, slow undetected hadrons were produced, appear at
$I>0$. For the cut $-0.05 < I < 0.05$
defining the exclusive sample the amount of the residual non-exclusive
background for the SMC experiment was up to about 10\% at large $Q^2$.
The kinematical smearing in $I$ and the
width of the elastic peak in COMPASS is expected to be about the
same (cf. Sect. 3.2) as that shown in Fig. \ref{ange} for the
SMC experiment. Although the smearing will be similar, we expect the
level of the non-exclusive background to be lower in COMPASS
due to the wider angular and momentum
acceptance coverage for final state hadrons. In addition,
with larger statistic in COMPASS it will be possible to
apply more tight inelasticity cuts, further reducing the background.
The effect of this residual background on various observables will
be studied by varying the inelasticity cuts, similarly as was done
in Ref. \cite{nmc1}.

The selections of coherent or incoherent production will
be done on a statistical basis, using the $t$-distribution;
at the lowest $\mid \! t \! \mid $ values coherent events predominate, whereas
at somewhat larger $\mid \! t \! \mid $ there is almost clean sample of incoherent
events.

Separation of the $\rho^{0}$ samples with the enhanced content
of longitudinally or transversely polarized mesons will be done by applying 
cuts on the measured angular distributions of pions from the decays
of the parent $\rho^{0}$.

The minimal covered $Q^2$ range is 
expected to be $0.05 < Q^2 < 10 \:\rm{GeV}^2 \!$.
\,For the \mbox{med-} \linebreak ium and large $Q^2$ values $(Q^2 > 2\:\rm{GeV}^2)$
the range $0.006 < x < 0.1$ will be covered with good acceptance.

The basic observable for each process studied
will be the ratio of the nuclear transparencies
for lead and carbon, $R^{}_{\rm T} = T_{\rm{Pb}}/T_{\rm{C}} =
(\sigma _{\rm{Pb}}/A_{\rm{Pb}})/(\sigma _{\rm{C}}/A_{\rm{C}})$. 
Due to the proposed geometry of the targets, the acceptances will 
cancel in the ratio $R^{}_{\rm T}$.
Also the absolute beam flux measurement
will not be necessary for the ratio $R^{}_{\rm T}$, provided that
the relative determination of the beam fluxes for the exposures with
different target materials could be done, e.g. by counting DIS events
originating in the polarized target.
The measured ratio $R^{}_{\rm T}$ should be corrected for
different losses of events in lead and carbon, which are due to 
the secondary interactions in the targets.
They will be estimated from the MC simulations.

\section{Simulation of exclusive $\rho^{0}$ events}

\label{lab_sec_3}

\hspace*{8mm}In this section we describe details of the simulation of exclusive
coherent $(\mu A \rightarrow \mu \, \rho^{0} A)$ and incoherent
$(\mu A \rightarrow \mu \, \rho^{0} N (A-1))$ $\rho ^0$ production in the COMPASS
experiment with the carbon and lead targets.
The simulations were done with a dedicated fast Monte Carlo
program which generates deep inelastic exclusive $\rho^{0}$
events with subsequent decay $\rho ^0 \rightarrow \pi^{+} \pi^{-} \!$.
\,At this stage there was no attempt to include any background in the
event generators.  
Here we describe the event generator as well as the treatment of
different experimental aspects: losses due to the  
secondary interactions of pions,
propagation through the magnetic fields,
angular and momentum resolutions, muon trigger acceptance, acceptance for 
final state pions and efficiency of tracks reconstruction.

Radiative corrections for exclusive $\rho ^0$ production 
are expected to be similar to those for the NMC experiment,
which were in the range of 2\% to 5\% \cite{kurek}.
These corrections were not included in the present simulations.
 
\subsection{Event generator}

\hspace*{8mm}First we present the used parameterization of the cross section for the 
production on the free nucleon, $\mu \, N \rightarrow \mu \, \rho^{0} N$,
with the subsequent decay $\rho^{0} \rightarrow \pi^{+} \pi^{-}$:
\begin{equation}
\sigma _{\mu N \rightarrow \mu \rho^{0} N} =
  \Gamma^{}_{\! T}(Q^2 \! , \: \nu) \cdot
  \sigma^{\rm tot}_{\gamma^{\ast} N \rightarrow \rho^{0} N}(Q^2 \! , \: \nu) \cdot
  F(p^{2}_{t}, \: \cos \theta, \: \phi ) \; ,
\end{equation}
where $\theta\/$ and $\phi\/$ are, respectively, the polar and azimuthal angles
of $\pi^{+}\/$ from the decay, calculated in the parent $\rho^{0}$
center-of-mass system, with respect to the direction of flight of $\rho^{0} \!$,
\,$\Gamma^{}_{\! T}\/$ is the flux of transverse virtual photons
\begin{equation}
\Gamma^{}_{\! T} = \frac{\alpha (\nu -\frac{Q^2}{2M^{}_{\! p}})}{2\pi Q^2 E^{2}_{\! \mu }
(1 - \epsilon)}\; ,
\end{equation}
$\alpha\/$ is the fine-structure constant,
$E^{}_{\! \mu}\/$ the muon-beam energy
and $\epsilon\/$ is the virtual
photon polarization given by
\begin{equation}
  \epsilon = \frac{1 - \frac{\nu }{E_{\! \mu }} - \frac{Q^2}{4E^2_{\! \mu }}}
                  {1 - \frac{\nu }{E_{\! \mu }}
                  + \frac{1}{2}(\frac{\nu }{E_{\! \mu }})^2
                  + \frac{Q^2}{4E^2_{\! \mu }}} \; .
\end{equation}

The virtual photon cross section was parametrized as
\begin{equation}
\sigma^{\rm tot}_{\gamma^{\ast} N \rightarrow \rho^{0} N}(Q^2 \! , \: \nu) = 27.4 \:\rm{nb}
\cdot \mbox{\Huge (} \frac{6\: \rm{GeV}^2}{Q^2} \mbox{\Huge )}^{\! \! 1.96}  .
\end{equation}
This is the NMC parametrisation of the 
cross sections per nucleon for exclusive $\rho^{0}$
production on carbon ~\cite{nmc}. As the NMC data shows
little $A$-dependence of the virtual photon \mbox{($t$-in-} \linebreak tegrated)
cross section per nucleon, we use
the same parametrisation of $\sigma^{\rm tot}_{\gamma^{\ast} N \rightarrow \rho^{0} N}$
for carbon and proton targets. This parametrisation does not 
apply at small $Q^2$ values, namely at $Q^2 < 1\: \rm{GeV}^2\!$.

The function $F\/$ comprises the $p^{2}_{t}\/$ distributions of produced $\rho^{0}$
and the angular distributions of pions coming from its decay
\begin{equation}
F = a^{}_{L} \cdot f^{}_{L} \, (p^{2}_{t}) \cdot W^{}_{\! L} \, (\cos \theta, \: \phi ) +
  a^{}_{T} \cdot f^{}_{\! T} \, (p^{2}_{t}) \cdot W^{}_{\! T} \, (\cos \theta, \: \phi ) \; .
\end{equation}
Here $a^{}_{L} = r^{04}_{00}$, $a^{}_{T} = 1 - r^{04}_{00}$,
$r^{04}_{00} = r^{04}_{00} (Q^{2} \!, \: \nu )$
is the $\rho^{0}$ density matrix element,
which can be identified as a fraction of longitudinally polarized
(helicity = 0) $\rho^{0}$ mesons,
and the indices $L\/$ and $T\/$ refer to longitudinally and transversely
polarized $\rho^{0}$'s, respectively.
The fraction $r_{00}^{04}$
can be expressed  ~\cite{SW} by the ratio $R$
of the cross sections for exclusive production by
longitudinal and transverse virtual photons.
We use a parametrisation of $R$ given by ~\cite{schildknecht}
which reproduce data on exclusive $\rho ^0$ production
in a wide range of $Q^2 \!$.

The $p^{2}_{t}\/$ distributions are described by 
\begin{equation}
f^{}_{\! i} (p^{2}_{t}) = b^{}_{i} \: e^{ -b^{}_{i} \:  p^{2}_{t}} \: ,
\end{equation}
where $i = L\/$ or $T\/$, $b^{}_{L} = 4.5 + 4 \cdot (0.5/Q^2) \: \rm{GeV}^{-2}$ and
$b^{}_{T} = 8.5 \: \rm{GeV}^{-2} \!$.
\,These parameterizations weighted by the fractions of longitudinally and 
transversely polarized mesons allow to reproduce reasonably the values of
the effective slope $b$ for the exclusive $\rho ^0$ production
measured at HERA and in the fixed-target experiments in a wide range of $Q^2 \!$.

The angular distributions of the pions from $\rho^{0}$'s decays are given by
\begin{equation}
 W^{}_{\! i} \, (\cos \theta, \: \phi ) = \frac{1}{2\pi } \: \frac{3}{4} \:
  \mbox{\Large \{} (1 - P^{}_{\! i}) \: + \:
(3 P^{}_{\! i} - 1) \cos^{2} \! \theta \mbox{\Large \}} \; ,
\end{equation}
where $i = L\/$ or $T\/$, $P^{}_{\! L} = 0$ and $P^{}_{\! T} = 1$.
For the nuclear targets we assume the same $W^{}_{\! i}$ distributions 
as for the proton as suggested by \cite{nmc}.
Note that the distributions $f^{}_{\! i} \, (p^{2}_{t})$, $W^{}_{\! i} \, (\cos
\theta, \: \phi )$ and $F(p^{2}_{t}, \: \cos \theta, \: \phi )$
are normalized to unity.

The invariant mass of two decay pions was generated using
the relativistic p-wave Breit-Wigner shape for the $\rho $
resonance \cite{jackson}.

We relate the differential cross sections for the proton
\begin{equation}
  \mbox{\Huge (} \frac{{\rm{d}} \sigma^{}_{\! N}}{{\rm{d}} t}
   \mbox{\Huge )}^{}_{\! \! i}
   \equiv \Gamma^{}_{\! T} \cdot \sigma^{\rm tot}_{\gamma^{\ast} N \rightarrow
   \rho^{0} N} \cdot a^{}_{i} \cdot f^{}_{\! i}(-t) \; ,
\end{equation}
to these for coherent and incoherent production on the nucleus $A$ by
\begin{equation}
 \mbox{\Huge (} \frac{{\rm{d}} \sigma^{\rm coh}_{\! \! A}}{{\rm{d}} t}
  \mbox{\Huge )}_{\! \! i} =
  A^{2}_{\rm eff \!, \; coh} \cdot e^{< R^{2}_{\! A} > \, t/3} \cdot
  \mbox{\Huge (} \frac{{\rm{d}} \sigma^{}_{\! N}}{{\rm{d}} t} \mbox{\Huge
  )}^{}_{\! \! i} \;\: ,
\end{equation}
\begin{equation}
 \mbox{\Huge (} \frac{{\rm{d}} \sigma^{\rm inc}_{\! \! A}}{{\rm{d}} t}
  \mbox{\Huge )}^{}_{\! \! i} =
  A^{}_{\rm eff \!, \; inc} \cdot \mbox{\Huge (} \frac{{\rm{d}} \sigma^{}_{\! N}}{{\rm{d}} t}
  \mbox{\Huge )}^{}_{\! \! i} \;\: ,
\end{equation}
respectively. Here $< \! R^{2}_{\! A} \! >$ is the mean squared radius of 
the nucleus,
$A^{}_{\rm eff \!, \; coh}$ and $A^{}_{\rm eff \!, \; inc}$ take account of nuclear screening
for the coherent and incoherent processes, correspondingly.
The cross section for incoherent
exclusive meson production is 
summed over all final states of the recoiling system,
i.e. it is given for the so called closure approximation. 
The suppression of the 
incoherent 
cross section at small $t$ due to the Pauli blocking is neglected here.
We used the approximation $t - t^{}_{\rm min} \simeq - p^{2}_{t}\/$, where
$ \mid \! t^{}_{\rm min} \! \mid$ is the minimal kinematically allowed $\mid \!
t \! \mid$ value for given $W^{2} \!$, \,$Q^{2} \!$, \,$m^{}_{\! V}\/$ and
$M^{2}_{\! X\/}$.

We generated the cross sections for two models. For the complete color transparency
model (CT model) 
we used $A^{}_{\rm eff \!, \; coh} = A^{}_{\rm eff \!, \; inc} = A\/$. 
In another model we assumed a substantial nuclear absorption (NA model) and used
$A^{}_{\rm eff \!, \; coh} = A^{}_{\rm eff \!, \; inc} = A^{0.75 \!}$.
These could be compared to different predictions
of the vector meson dominance model (VMD), 
which vary depending on the $\rho $ meson-nucleon
total cross section and on the coherence length $l_c$
(see e.g. \cite{KM}). 
The $A$ dependences of the NA model are in the range of predictions
of VMD, except for incoherent production at large $l_c$, where VMD
predicts stronger nuclear absorption.

\subsection{Simulation of the experimental effects}

\hspace*{8mm}The secondary hadronic interactions of the decay pions
in the target were simulated. The assumed
density of targets, $\rho^{}_{\rm tgt}\/$, was 2.2 g/cm$^3 \!$.
\,The interaction length $\lambda^{\pi}_{\rm int}\/$ for pions
in the target material, was assumed equal to 130 g/cm$^2$ for the
carbon and 290 g/cm$^2$ for the lead target.
For an exclusive $\rho ^0$ event to be reconstructible,
it was required that none of the decay pions 
underwent an inelastic hadronic interaction. 

The trajectories of charged particles were simulated taking into account
the geometry of SM1 and SM2 magnets of the COMPASS experiment.
Homogeneous fields inside both magnets were assumed. For the SM1 magnet
the bending power
$\int \! B \, {\rm{d}}l\/$ was assumed equal to 1.0 T$\cdot$m (independent of
the beam energy), whereas for the SM2 magnet it was assumed equal
to 2.3 T$\cdot$m for a 100 GeV muon beam energy,
and 5.2 T$\cdot$m for a 190 GeV muon beam.
After tracking of the produced charged particles through the detector,
a flag was assigned to each particle telling how far it propagated
in the COMPASS setup.

Kinematic smearing of the beam, of the scattered muon and of the charged hadrons
was simulated. Assumed values of dispersions of
measured particle momenta and angles are based on the experience
of previous muon experiments at CERN, as well as on the
results of studies at COMPASS. The relative error on the momentum, $\sigma (p)/p$, was assumed
equal to 0.5\% for beam tracks, 0.75\% for the tracks
passing only the first magnet and 0.44\% for the tracks passing the second
magnet. The error on the angle of a particle was assumed to be 0.15 mrad.
For a 190 GeV beam and the kinematic cuts listed in Section \ref{lab_sec_3}
the resulting smearing of the inelasticity $I$ is about 0.018 and the smearing
of the invariant mass of two pions is about 6 MeV.

To simulate the trigger acceptance a trajectory of the scattered muon behind the
second magnet was calculated, and the hits in the muon hodoscopes H4 and
H5 were checked. Each of these two hodoscopes consists in fact of
a few different hodoscopes (namely of the Ladders, the Primed System, the Unprimed
System) but for the purpose of the present analysis we will consider just two
cases, corresponding to different trigger acceptances.
First, we assumed that only the Ladders and the Primed System
are available. This trigger is called the Medium $Q^{2}$ range Trigger (MT).
If in addition the Unprimed System is also implemented, the Full $Q^2$ range
Trigger (FT) will result.
The $Q^2$ dependence of the trigger efficiency, $\epsilon^{}_{\rm tr}\/$,
for these triggers was presented in \cite{memo2000}.
For the MT trigger at 190 GeV beam $\epsilon^{}_{\rm tr}$ decreases quickly with $Q^2$
from 0.7 at $Q^2 = 2 \:\rm{GeV}^2$ to about 0.1 at $Q^2 = 10 \:\rm{GeV}^2 \!$.
\,The acceptance is several times smaller for this trigger at 100 GeV beam.
For the FT trigger the $Q^2$ dependence is weaker and the trigger acceptance
is higher; it is always bigger than 0.5 for  $Q^2 < 70 \:\rm{GeV}^2$ at
190 GeV beam energy and for $Q^2 < 20 \:\rm{GeV}^2$ at 100 GeV beam.
In conclusion, for the MT trigger the data taking with
the higher beam energy seems the only acceptable choice, whereas
for the FT trigger running at both beam energies
is feasible, although the covered $Q^2$ range is larger at
the higher beam energy. 

For a $\rho^0$ meson to be accepted it was required that each pion from its
decay was emitted in the laboratory at an angle within the acceptance of
the SM1 magnet, and that its momentum was bigger than 2$\;$GeV.

Based on the preliminary results of the track reconstruction by the
programs developed at COMPASS, simple and rather conservative
assumptions were used to simulate
the efficiency of tracks reconstruction. For the tracks seen only in the
first spectrometer the single track reconstruction efficiency was assumed equal
to 0.8 for the momentum range $p > 2 \:\rm{GeV}$, and for the tracks
observed also in the second spectrometer it was assumed equal to 0.95.
The efficiency $\epsilon^{}_{\rm rec}\/$ to reconstruct all three tracks
of the scattered muon and of two pions
was assumed to be equal $0.7 \cdot 0.95^3 + 0.3 \cdot 0.95^2 \cdot 0.80$.
This assumption 
was motivated by the observation that for 70\% of accepted DIS exclusive
$\rho^{0}$ events all three measured tracks are seen in the second spectrometer,
while for the remaining 30\% events the scattered muon and the fast pion are
seen in both first and second spectrometers, whereas the slow pion is observed
in the first spectrometer only.

\section{Results on exclusive $\rho ^0$ production}

\label{lab_sec_4}

\hspace*{8mm}Due to the higher trigger efficiency and larger $Q^2$ range at higher
beam energy we considered 190 GeV muon beam. The simulations were 
done independently for the carbon $(A = 12)$ and lead targets $(A = 207)$,
and for two triggers (MT and FT).
For each target and each trigger we assumed two different
models describing the nuclear effects for exclusive $\rho^{0}$ production:
CT model and NA model (cf. Section 3.1).

The kinematic range considered was the following:
\begin{equation}
\label{EQ_MC5} 2 < Q^2 < 80\: \rm{GeV}^2 \: ,
\end{equation}
\begin{equation}
\label{EQ_MC6}
35 < \nu < 170\: \rm{GeV} \: .
\end{equation}
The upper cut on $\nu $ was chosen to eliminate the kinematic
region where the amount of radiative events is large,
whereas the lower one to eliminate the region where the acceptance
for pions from $\rho^{0}$ decay is low.

The total efficiency $\epsilon^{}_{\rm tot}\/$ to observe exclusive $\rho^{0}$ events
results from: the acceptance of the trigger $(\epsilon^{}_{\rm tr})\/$,
acceptance to detect the pions $(\epsilon^{}_{\rm had})\/$,
efficiency for tracks reconstruction $(\epsilon^{}_{\rm rec})\/$,
cut on the invariant mass of two pions, $0.62 < M^{}_{\pi \pi } < 0.92 \: \rm{GeV}^2 \!$,
\,used for the selection of the samples ($\epsilon^{}_{\rm mass}\/$),
and efficiency for an event to survive the 
secondary interactions $(\epsilon^{}_{\rm sec})\/$.
The contributions of all these effects to $\epsilon^{}_{\rm tot}\/$ are similar to
those presented in \cite{memo2000} for the polarized target,
except for the effects of the secondary interactions.
The approximate value of $\epsilon^{}_{\rm sec}\/$ is equal to
0.87
for the carbon target and
0.94
for the lead target. The total efficiency $\epsilon^{}_{\rm tot}\/$ is about
0.48
for the FT trigger and about
0.30
for the MT trigger. 

The total expected cross section for exclusive $\rho ^0$ production on the nucleon is
\begin{equation}
 \sigma^{\rm tot}_{\mu N \rightarrow \mu \rho^{0} N} = \int_{\nu^{}_{\rm min}}^{\nu^{}_{\rm max}}
  \! \! \int_{Q^{2}_{\rm min}}^{Q^{2}_{\rm max}}
  \! \! \sigma^{}_{\mu N \rightarrow \mu \rho^{0} N} (Q^{2} \!, \: \nu ) \:
  {\rm d} Q^{\! 2} \, {\rm d} \nu \; , 
\end{equation}
where the kinematic range was defined before.
The value of $\sigma^{\rm tot}_{\mu N \rightarrow \mu \rho^{0} N}\/$ is 283 pb.
For nuclear targets the corresponding values depend on $A$ and on 
the assumed model
for nuclear absorption. 

The expected muon beam intensity will be about $10^{8} \! / \! \rm{s}$ during
spills of length of about 2$\:$s, which will repeat every 14.4$\:$s. With the proposed
thin nuclear targets, each of 17.6 g/cm$^2$, the luminosity will be
${\cal{L}} = 12.6 \:\rm{pb}^{-1} \cdot \rm{day}^{-1} \!$.
\,The estimates of the numbers of accepted events were done for a
period of data taking of 150 days (1 year),
divided equally between two targets. An overall SPS and COMPASS
efficiency of 25\% was assumed. The numbers of accepted events for
the carbon and lead targets, assuming the two models
for the nuclear absorption mentioned earlier, are given in Table \ref{rates}.


\begin{table}[t]

\vspace*{-5mm}

\caption{Numbers of accepted events for two considered models
	 of the nuclear absorption.
\label{rates}}
\begin{center}
\footnotesize
\begin{tabular}{|c|c|c|}
\hline
\hline
  \raisebox{0mm}[5mm][3mm]{\hspace*{3mm} {\large model} \hspace*{3mm}}  &  {\large \hspace*{6mm} $N^{}_{\rm C}$ \hspace*{6mm}}
   &  {\large \hspace*{6mm} $N^{}_{\rm Pb}$ \hspace*{6mm}}  \\
\hline
\hline
  \raisebox{0mm}[5mm][3mm]{\large CT}   &  {\large 70 000}  &  {\large 200 000}  \\
\hline
  \raisebox{0mm}[5mm][3mm]{\large NA}   &  {\large 28 000}  &  {\large 20 000}   \\
\hline
\hline
\end{tabular}
\end{center}
\end{table}


The distributions of accepted exclusive events
as a function of $x$, $Q^2$ and $\nu $
are similar to those presented in \cite{memo2000}.

\begin{figure}[t]
 
\begin{center}
\mbox{\epsfig{file=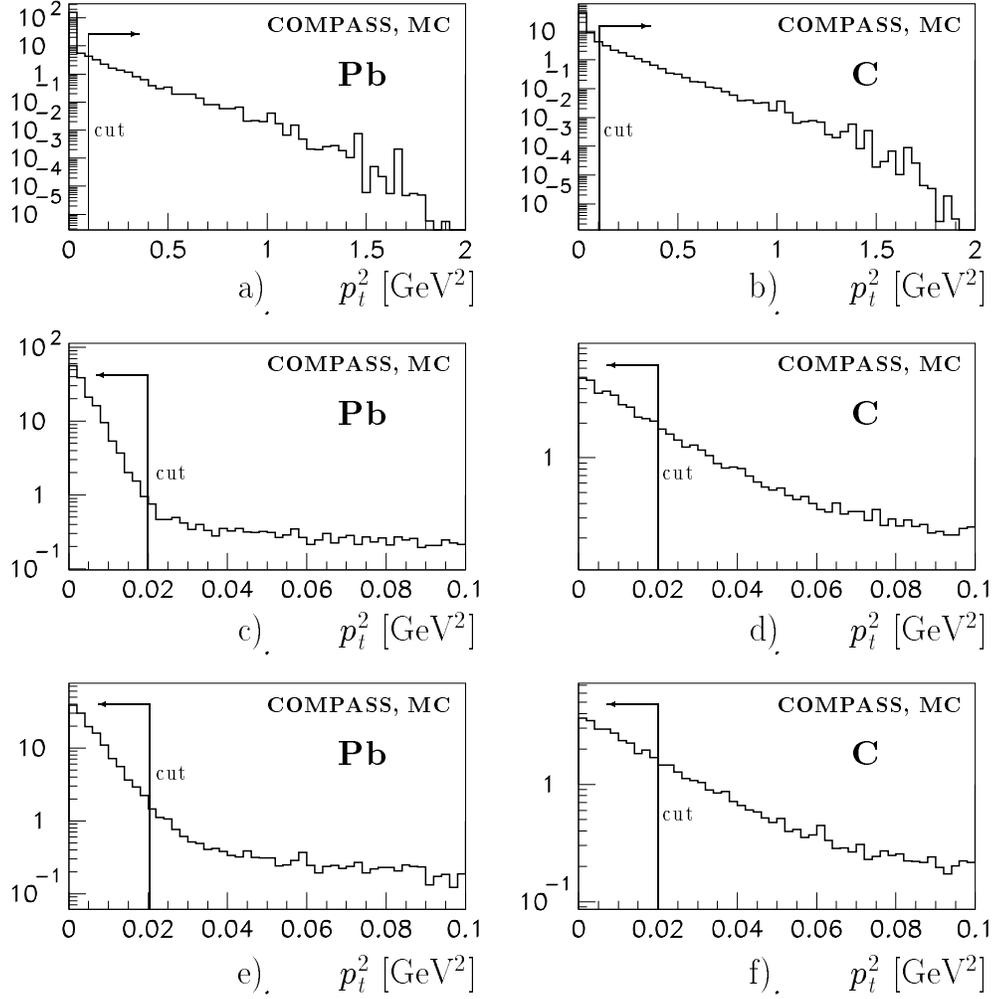,height=15cm}}
\end{center}

 \caption{Distributions of $p^{2}_{t}\/$ for the lead (left) and carbon
  (right) targets. The distributions a) and b) at the top are for the generated
  events and the wide range of $p^{2}_{t}\/$, whereas those in the middle,
  c) and d),
  correspond to the range of low $p^{2}_{t}\/$. The bottom distributions,
  e) and f), are for the accepted events, with the kinematical smearing
  taken into account.
  The arrows show the cuts described in the text. \label{pt2}}
 \end{figure}

In Fig.$\;$\ref{pt2} we present the $p^{2}_{t}\/$ distributions for both
targets. We observe clear coherent peaks at small $p^{2}_{t}\/$
($< 0.05 \:\rm{GeV}^2$) and less steep distributions for
the incoherent events at larger $p^{2}_{t}\/$. The
arrows
at the top histograms indicate the cut $p^{2}_{t} > 0.1 \:\rm{GeV}^2 \!$,
\,used to select the incoherent samples. The contribution from coherent
events is negligible in these samples. For the middle and bottom histograms
the arrows
indicate the cut $p^{2}_{t} < 0.02 \:\rm{GeV}^2 \!$, \,used to select the
samples which are dominated by coherent events --- the \linebreak so called coherent samples.
For the samples defined by the latter cut
the fraction of
the incoherent events is at the level of up to 10\%, depending on the nucleus
and on the model for nuclear absorption.
We plan also to use the standard 
method to determine coherent and incoherent
components, by fitting $p_t^2$ distribution. 

The effect of the kinematical smearing on $p^{2}_{t}\/$ may be
seen by comparing the distributions for the generated events
(middle row) to the ones for measured events (bottom raw) where the 
acceptance and smearing were included. The smearing of $p^{2}_{t}\/$
increases with increasing $p^{2}_{t}\/$; it is about $0.006 \:\rm{GeV}^2$
for the coherent samples ($p^{2}_{t} < 0.02 \:\rm{GeV}^2$)
and about $0.03 \:\rm{GeV}^2$ for the
incoherent samples ($p^{2}_{t} > 0.1 \:\rm{GeV}^2$).

\begin{figure}[t]

\begin{center}
\mbox{\epsfig{file=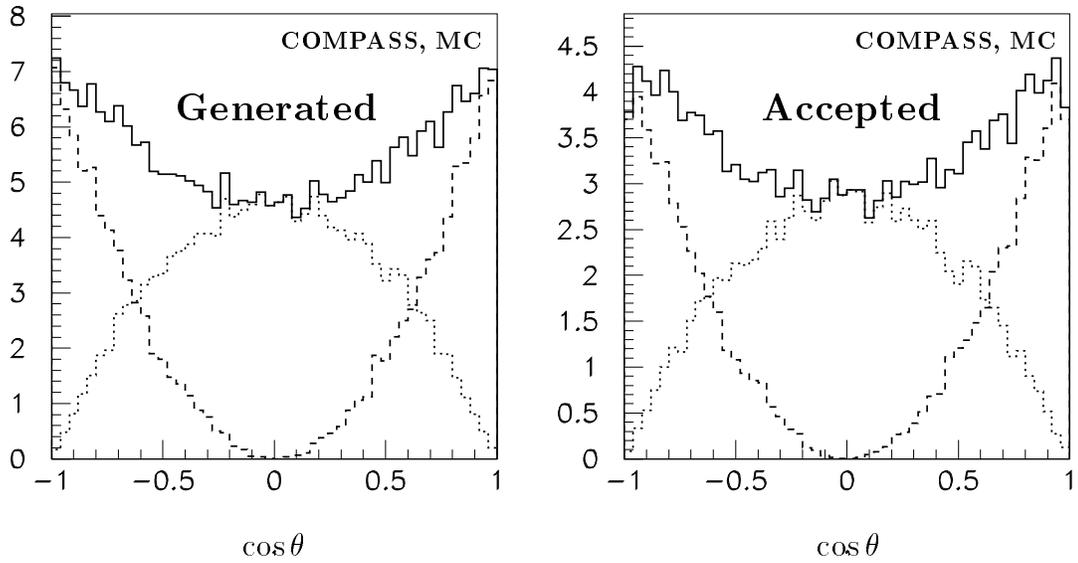,height=10cm}}
\end{center}

 \caption{Distributions of $\cos \theta \/$ 
  for the pions $(\pi^{+})$ from the decays of 
  $\rho^{0}\/$ mesons produced on lead, for generated (left)
  and accepted (right) events. For the latter ones the effects
  of the kinematical smearing were taken into account.
  The dashed- and dotted-line histograms are for the longitudinally 
  and transversely polarized $\rho^{0}\/$'s, respectively, and the solid-line
  histograms are for their sum. \label{costh}}
 \end{figure}

The analysis of the $\rho $ decay distributions allows us to study
spin-dependent properties of the production process \cite{SW},
in particular the polarization of $\rho $.
Usually the $\rho^{0}$ decay angular distribution
$W(\cos \theta, \: \phi )$
is studied in the $s$-channel helicity frame, which is the most
convenient for describing the $\rho $ decay after photo- and
electroproduction \cite{angdis}. The $\rho ^0$ direction in the virtual
photon-nucleon
centre-of-mass system is taken as the quantization axis. The angle
$\theta \/$ is the polar angle and $\phi\/$ the azimuthal angle
of the $\pi^{+}\/$ in the $\rho^{0}$ centre-of-mass system.
The $\cos \theta \/$ distributions for pions from $\rho^{0}$ decays
are shown in Fig.$\;$\ref{costh} for the lead target. 
The distributions for longitudinally (dashed lines) and transversely
(dotted lines) polarized parent $\rho^{0}$'s are markedly different. 
Their sum is also indicated.

Usually fits to the combined $\cos \theta \/$ distributions
are performed in order to determine the density matrix element $r^{04}_{00}$
(cf. Eq. 11), which can be identified as the probability
that the $\rho ^0$ was longitudinally polarized.
For exclusive
$\rho^{0}$ production the
approximate $s$-channel helicity conservation (SCHC) is observed
\cite{zeus1,h1}, i.e.
the helicity of $\rho^{0}$ is predominantly equal to that of the
virtual photon.
Assuming SCHC and using the fitted $r^{04}_{00}$ one can estimate the 
ratio $R = \sigma _L/\sigma _T$ for exclusive virtual photoproduction
(\cite {SW}).
In COMPASS we plan to measure $R$ as well, and study its $Q^2$- and
$A$-dependence,
which is expected to reflect
the strength of nuclear absorption.

As $Q^2$ increases, the approach to the CT limit is expected to be different
for VMP by longitudinally polarized virtual photons from that by
transversely polarized photons. Therefore, in order to increase the
sensitivity of the search for CT, we propose another method.
It consists in studying $A$-dependence
of the cross sections for samples with different $\rho ^0$ polarizations,
which will be
selected by cuts on $\cos \theta \/$.
For instance,
after applying the cut
$\mid \! \cos \theta  \! \mid > 0.7$ the fraction of accepted events
with $\rho^{0}_{L}\/$ is
(80--95)\% depending on the simulation,
whereas for the cut $\mid \! \cos \theta  \! \mid < 0.4$ the
fraction of events with $\rho^{0}_{T}\/$ is (75--92)\%. 
For an approximate SCHC,  
such cuts will allow us to select the samples
with enhanced contributions of the events
initiated by the virtual photons of a desired polarization.
Studies of the samples with {\it different polarizations\/} of the 
virtual photons are {\it important\/} for the clear demonstration of CT.

\begin{figure}[t]
\begin{center}
\mbox{\epsfig{file=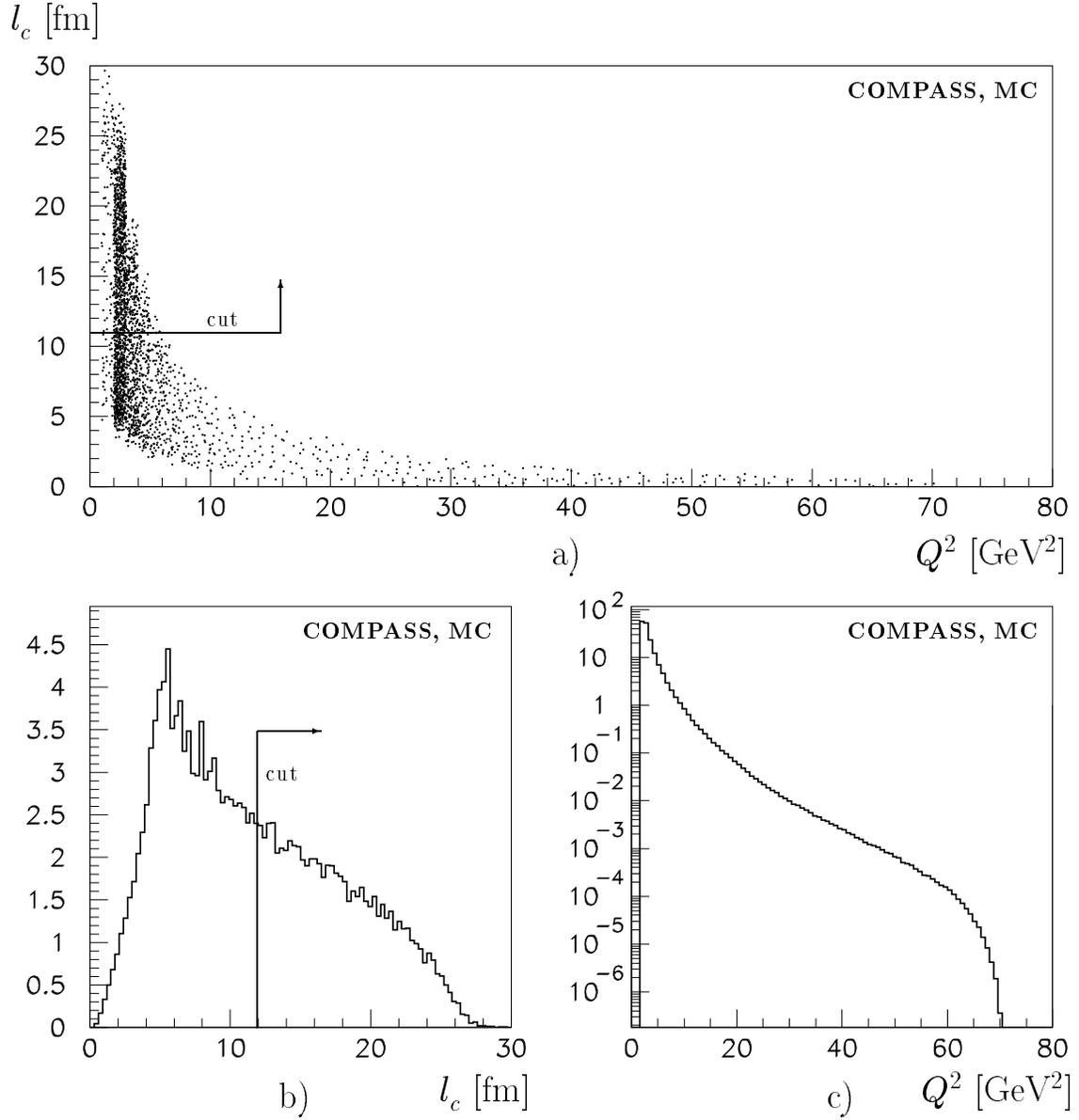,height=18cm}}
\end{center}

 \caption{Correlation of $l^{}_{c}\/$ and $Q^{2}\!$. \,The plot a) and its 
  projections b) and c) were obtained for the accepted events from $\rho^{0}\/$ 
production on lead. 
  The arrows show the cut discussed in the text. \label{lcq2}}
 \end{figure}

Another aspect which is important for CT studies, is the covered range
of the coherence length $l^{}_{c}\/$ (cf. Section 1). 
In Fig. \ref{lcq2} we
present the plot of $l^{}_{c}\/$ vs. $Q^{2}\/$ for a sample of accepted events.
The effects of initial and final state interactions in the nuclei
vary at small $l_c$ values \cite{HERMES}. Therefore, it was suggested 
that in order to disentangle effects due to CT from those caused by 
the modified absorption at small $l_c$ values, one should study
$A$- and $Q^2$-dependences of cross sections at fixed values of $l_c$.
This approach will be possible, if large statistics data were available.

For a limited statistics, a possible solution to avoid the mentioned
effects is to use the combined data at $l_c$ values exceeding the sizes
if the target nuclei.
The radius of the carbon nucleus
is $< r^{2}_{\rm C} >^{1/2} = 2.5\:\rm{fm}$ and that of the lead nucleus is
$< r^{2}_{\rm Pb} >^{1/2} = 5.5\:\rm{fm}$ \cite{radii}.
Therefore, 
one may use the selection
$l^{}_{c} > l^{\rm min}_{c} \simeq 2 \cdot < r^{2}_{\rm Pb} >^{1/2} = 11 \:\rm{fm}$.
The value of $l^{\rm min}_{c}\/$ is indicated in Fig.$\;$\ref{lcq2} by
arrows.
About a half of events survive the cut on $l^{}_{c}$. These events
cover the range of $Q^{2} < 6 \:\rm{GeV}^2 \!$, \,which is expected to be
sufficient to observe CT. 

\begin{figure}[t]

\begin{center}
\mbox{\epsfig{file=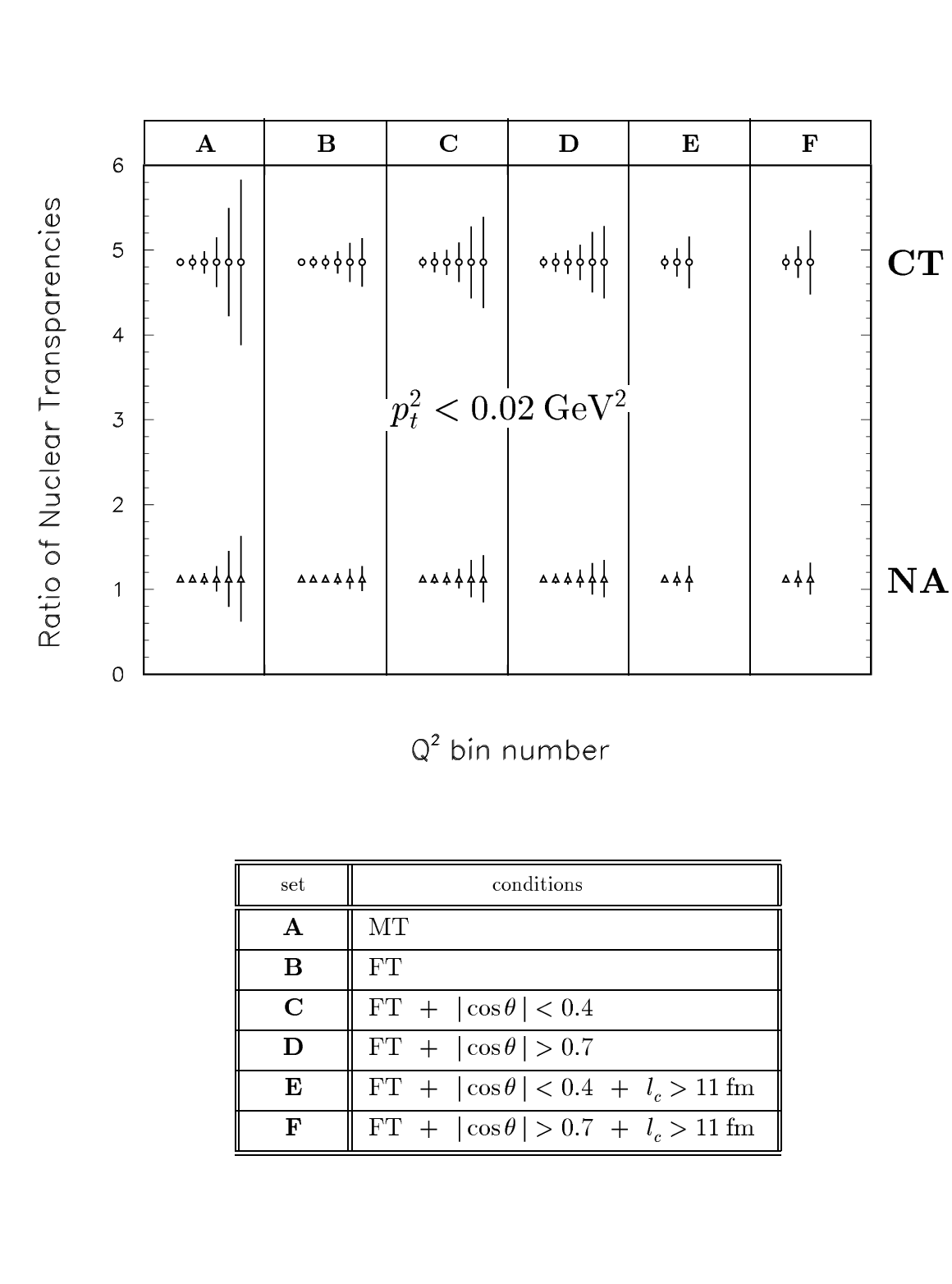,height=18cm}}
\end{center}

 \caption{Expected statistical precision of the
  measurement of the ratio
  $R^{}_{\rm{T}}\/$ of
  nuclear transparencies for lead and carbon for $p^{2}_{t} < 0.02 \:\rm{GeV}^2$
  and for different $Q^{2}$ bins, for CT (upper band) and for NA (lower band) models.
  The $Q^{2}$ bins are defined in Table 3. Sets A, B, C, D, E 
  and F correspond to the
  conditions specified in the table below the plot. \label{ratcoh}}
 \end{figure}

\begin{figure}[t]

\begin{center}
\mbox{\epsfig{file=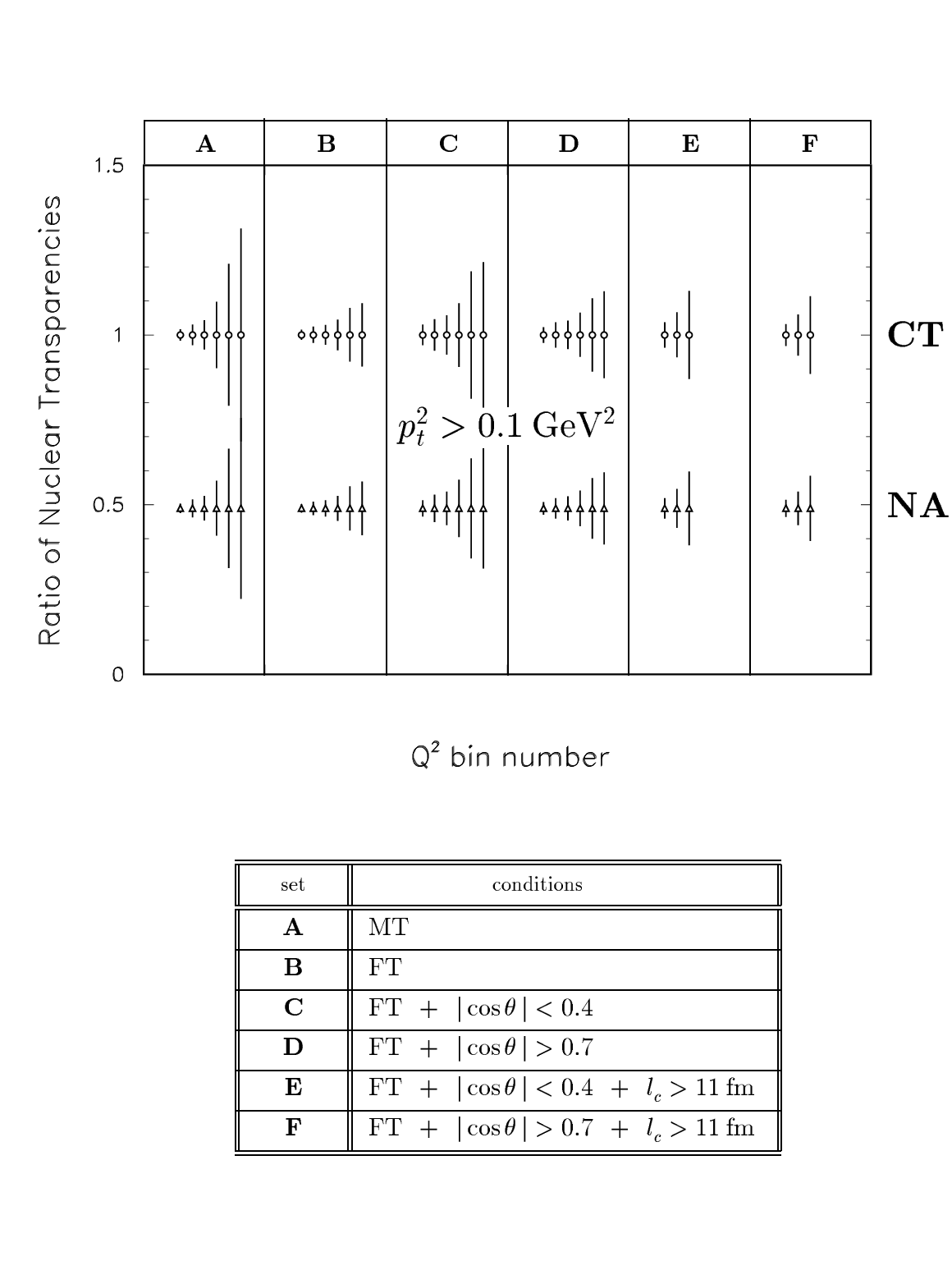,height=18cm}}
\end{center}

 \caption{Expected statistical precision of the
  measurement of the ratio
  $R^{}_{\rm{T}}\/$ of
  nuclear transparencies for lead and carbon for $p^{2}_{t}\/ > 0.1 \:\rm{GeV}^2$
  and for different $Q^{2}$ bins, for CT (upper band) and for NA (lower band) models.
  The $Q^{2}$ bins are defined in Table 3. Sets A, B, C, D, E 
  and F  correspond to the
  conditions specified in the table below the plot. \label{ratinc}}
 \end{figure}

The estimated values and 
statistical precision of $R^{}_{\rm T}$, the ratio of the nuclear transparencies
for lead and carbon, are presented for different $Q^{2}$ bins
in Fig.$\;$\ref{ratcoh} for $p^{2}_{t} < 0.02 \:\rm{GeV}^{2}$ and
in Fig.$\;$\ref{ratinc} for $p^{2}_{t} > 0.1 \:\rm{GeV}^{2} \!$.
\,The $Q^{2}$ bins are specified in Table \ref{q2bins}.
Each figure comprises predictions for two models, CT and NA,
and for 6 different samples of accepted events for each model.
For each sample a set of "measurements" in different $Q^2$ bins is shown.
Sets {\bf A} and {\bf B} were obtained using the standard selections for
the MT and FT triggers, respectively. Four remaining sets
were obtained for the FT trigger with additional selections:
{\bf C} with $\mid \! \cos \theta \! \mid < 0.4$, {\bf D} with $\mid \! \cos
\theta \! \mid > 0.7$,
{\bf E} with $\mid \! \cos \theta \! \mid < 0.4$ and $l^{}_{c} > 11\:\rm{fm}$, and
{\bf F} with $\mid \! \cos \theta \! \mid > 0.7$ and $l^{}_{c} > 11 \:\rm{fm}$.
Note that for sets {\bf E} and {\bf F} only three lower $Q^{2}$ bins
appear (cf. Fig.$\;$\ref{lcq2}).
One expects large differences in $R^{}_{\rm T}$ for the two considered models.
For coherent samples $R^{}_{\rm T} \approx 5$ for CT model and 
$\approx 1$ for NA model.
At $Q^2 \simeq 5 \:\rm{GeV}^2$ the precision of the measurement of
$R^{}_{\rm T}$ for coherent events will be better than 17\%, even for the
restricted samples {\bf E} and {\bf F}, thus allowing excellent discrimination
between the two models for the nuclear absorption.
For the incoherent events the power to discriminate models by 
$R^{}_{\rm T}$ measurements
will be more limited.


\begin{table}[t]

\vspace*{-5mm}

\caption{$Q^{2}$ bins used for the determination of $R^{}_{\rm T}$.
\label{q2bins}}
\begin{center}
\begin{large}
\begin{tabular}{|c|c|c|c|}
\hline
\hline
 \raisebox{0mm}[5mm][2mm]{Bin number}  &  \ \ \ \ $Q^{2}$ bin\ \ \ \   &
  			  \ \ \ \ $<Q^{2}>$\ \ \ \   &  \ \ \ \ $<x>$\ \ \ \   \\
   & \raisebox{0mm}[4mm][3mm]{$[\rm{GeV}^2]$} & $[\rm{GeV}^2]$ &\\
\hline
\hline
  \raisebox{0mm}[5mm][3mm]{1}  &  2--3  &  2.4  &  0.016  \\
\hline
  \raisebox{0mm}[5mm][3mm]{2}  &  3--4  &  3.4  &  0.022  \\
\hline
  \raisebox{0mm}[5mm][3mm]{3}  &  4--6  &  4.8  &  0.031  \\
\hline
  \raisebox{0mm}[5mm][3mm]{4}  &  6--9  &  7.2  &  0.048  \\
\hline
  \raisebox{0mm}[5mm][3mm]{5}  &  9--12  &  10.2  &  0.072  \\
\hline
  \raisebox{0mm}[5mm][3mm]{6}  &  12--20  &  14.8  &  0.11  \\
\hline
\hline
\end{tabular}
\end{large}
\end{center}
\end{table}


\section{Comparison with previous experiments}

\label{lab_sec_5}

\hspace*{8mm}In this section we compare COMPASS capabilities to
demonstrate CT to those of previous experiments in which exclusive
$\rho ^0$ leptoproduction on nuclear targets was studied. We concentrate on
this reaction, as among different exclusive VMD channels it has
the largest cross section. For the comparison we have selected
experiments which covered $Q^2$ range extending to large values,
bigger than 2 GeV$^2$. This condition is satisfied for the
following experiments: HERMES \cite{HERMES}, NMC \cite{nmc} and 
E665 \cite{E665}.

The published results from the HERMES experiment concern the incoherent
exclusive $\rho ^0$ production on $^1{\rm H}$, $^2{\rm H}$, $^3{\rm He}$
and $^{14}{\rm N}$  targets. The electron beam energy was 27.5 GeV
and the covered $Q^2$ and $l_c$ ranges are $0.4 < Q^2 < 5.5\: {\rm GeV}^2$
and $0.6 < l_c < 8$ fm.
Like in all fixed-target experiments with lepton beams,
the covered kinematic range is such that for a given $Q^2$
the average value of $l_c$ decreases with increasing $Q^2$.
The highest values of the HERMES data where one may expect to observe
the onset of CT are correlated with the small values of $l_c$
below 2 fm. The increase of the nuclear transparency observed
for $^{14}{\rm N}$ is explained as due to the reduced coherence length,
without resorting to CT.

Wider kinematic ranges were coverd by experiments using high-energy
muon beams. NMC has published the data on coherent and incoherent
exclusive $\rho ^0$ production on $^2{\rm H}$, C and Ca targets.
The muon beam energy was 200 GeV and the data cover the ranges
$2 < Q^2 < 25\: {\rm GeV}^2$ and $1 < l_c < 30$ fm.
For the NMC kinematic range, the large values of $l_c$
which exceed the diameter of the largest target nucleus (Ca),
$l_c > 7$ fm, correspond to $Q^2 < 5.5\: {\rm GeV}^2$.
In principle, in this range it is possible to study $Q^2$
dependence of nuclear absorption in order to observe CT
not obscured by effects of small $l_c$.
However, due to the moderate statistics of the data, such
detailed analysis was not possible for NMC.

The most favourable kinematical conditions were realised in the
E665 experiment due to the high muon beam energy of 470 GeV.
The experiment has published the data on incoherent
exclusive $\rho ^0$ production on $^1{\rm H}$, $^2{\rm H}$, C, Ca and Pb
targets. The data were taken in a wide $Q^2$ range, including
also very small values $Q^2 > 0.1\: {\rm GeV}^2$, and in a wide range
of $l_c$, $1 < l_c < 200$ fm. At small and moderate $Q^2$ values,
$< 3$ GeV$^2$, which correlate with $l_c$ values exceeding
the diameter of the lead nucleus, precise measurements of
the nuclear transparency were obtained. They indicate strong
nuclear absorption. At the highest $Q^2$ bin,
$Q^2 > 3\: {\rm GeV}^2$, the nuclear transparency increases in
qualitative agreement with CT. However, the data in this bin correspond 
to a wide range of $l_c$, $1 < l_c < 40$ fm, 
and therefore could be affected by effects due to the reduced
coherence length. The low statistics in the large $Q^2$ bin
did not allow more detailed studies.

The COMPASS data at medium and large $Q^2$ (FT trigger) will cover
the kinematic range similar to that of the NMC data.
The expected statistics for the carbon target will be
about 2 orders of magnitude higher than that of the NMC data
for the same target. This increase is in particular due to higher
beam intensity and larger acceptance in the COMPASS experiment.
Similarly like in the E665 experiment COMPASS will use also
lead target and will extend measurements to small $Q^2$.
The statistics of COMPASS data will be significantly
higher also than that of E665. For example, for incoherent events
at $Q^2 > 3\: {\rm GeV}^2$ the ratio of nuclear transparencies for
lead and carbon will be determined with accuracy better than 3\%,
comparing to about 30\% for E665.
Another more sensitive method to study nuclear effects in exclusive
$\rho ^0$ production, the one by using coherent production,
will be employed in COMPASS as well.
Due to large statistics splitting of COMPASS data in several $Q^2$
and $l_c$ bins as well as the selection of events with longitudinal
or transverese $\rho ^0$ polarization will be possible. 

\section{Open questions}

\label{lab_sec_6}

\hspace*{8mm}In addition to the $\rho ^0$ production, we will
study exclusive production of
$J \! / \! \psi\/$, $\phi $,
$\psi'$ and $\rho'$ and their ratios in order to
demonstrate CT. For these channels the analysis will be similar
to that for $\rho^{0} \!$, \,but the rates will be lower due to
the smaller production cross sections. The estimates of the
required period of data taking and of the target thicknesses
are subject of further analysis.

We consider also a possibility to study CT via coherent production
of hadron pairs with large relative transverse momenta.
The experimental procedure would be similar to that used in the
Fermilab experiment E791 \cite{Wei97}, in which CT was observed in
coherent diffraction of pions into high-$p_t$ di-jets.
We will select a sample of events with high-$p_t$ hadron pair 
and an additional requirement that all observed charged hadrons
carry at least 90\% of the energy of $\gamma ^*$.
An estimate of $t$ will be obtained using momenta of
all measured charged hadron tracks. The expected resolution
in $t$, which will be crucial for selection of the coherent sample,
will be determined from 
a dedicated simulation and reconstruction of events with
high-$p_t$ hadron pairs.

Finally we mention few experimental aspects which have to
be further investigated.\\ 

Here we propose to position the nuclear
targets downstream of the polarized target. Then the CT
physics could complement the standard COMPASS program 
by using the downstream nuclear targets
simultaneously with the polarized target.
Similar target configuration was already used during
the 1985 running of the EMC experiment, 
when the polarized target was used together
with several nuclear targets situated downstream of PT.

The effect of the downstream nuclear targets
on the hadrons coming from the polarized target should be
small. For instance, for the reconstructible 
tracks of hadrons from PT, only about 4\% of slow pions from
$D^{\ast}$ decays and about 1\% of hadrons from $D^{0}$ decays
will be in the acceptance of the proposed nuclear target
\cite{adam}. For the scattered muon coming from PT and passing through
the nuclear target, the additional multiple Coulomb scattering
should result in a small 
increase of the errors of the reconstructed position of the event vertex.

Assuming a dedicated run with unpolarized targets,
different target configurations for CT studies are possible.
With more available space
one could envisage targets
at least about 4 times thicker 
than the thin targets presented earlier.
Thus the four-fold reduction of the data taking time
will be possible. With an even larger amount of available beam time, one could
expose more nuclear targets of different materials, and in addition
the liquid hydrogen or deuterium targets. The latter ones are also important
for measurements of {\it off-forward parton distributions\/}
(also referred to as {\it skewed parton distributions\/} or
{\it generalized parton distributions\/}) \cite{off}.
For a dedicated run with unpolarized targets, one could take
advantage of a possibility of adding a recoil detector. 
This is crucial for certain processes, e.g.
Deep Virtual Compton Scattering (DVCS),
but in any case it will help to reduce the non-exclusive background
for any exclusive process.

\section{Conclusions}

\label{lab_sec_7}

\hspace*{8mm}We have simulated and analysed exclusive $\rho^{0}$ muoproduction
$(\mu A \rightarrow \mu \, \rho^{0} A$ and 
$\mu A \rightarrow \mu \, \rho^{0} N (A-1))$ at the COMPASS experiment
using thin nuclear targets of carbon and lead.
For muon beam energy of 190 GeV and a trigger
for medium and large $Q^{2}$, the covered kinematic range is
$2 < Q^2 < 20\: \rm{GeV}^2$ and $35 < \nu < 170\: \rm{GeV}$.

Good resolutions in $Q^{2} \!$, \,$l^{}_{c}$, $t$ ($p^{2}_{t}$) and $\cos
\theta \/$ are feasible.
An efficient selection of coherent or incoherent events is possible
by applying cuts on $p^{2}_{t}$. In order to obtain the samples of events
initiated with a probability of about 80\% by either $\gamma^{\ast}_{L}\/$
or $\gamma^{\ast}_{T}\/$, the cuts on the $\rho^{0}$ decay
angular distribution of $\cos \theta \/$ will be used.
The search for CT
could be facilitated by using the events with $l^{}_{c}\/$ values exceeding
the sizes of the target nuclei. The fraction of such events is
substantial and the covered $Q^{2}$ range seems sufficient to observe CT.

We showed high sensitivity of the measured ratio $R^{}_{\rm T}$ of nuclear 
transparencies
for lead and carbon for different models of nuclear absorption.
Good statistical accuracy of the measured $R^{}_{\rm T}$ 
may be achieved already for one year of data taking.
These measurements, taken at different $Q^{2}$ intervals, may allow to
discriminate between different mechanisms of the interaction of the hadronic
components of the virtual photon with the nucleus.

In conclusion, the planned comprehensive studies of exclusive vector meson
production on different nuclear targets at the COMPASS experiment
should unambiguously demonstrate CT.

\section{Acknowledgements}
\label{lab_sec_8}

\hspace*{8mm}The authors gratefully acknowledge useful
discussions with L. Frankfurt.
We thank also S. Kananov for discussions on the ZEUS exclusive
$\rho ^0$ data.
The work was supported in part by 
the Israel Science Foundation (ISF) founded by the Israel Academy of
Sciences and Humanities, Jerusalem, Israel and
by the Polish State Committee for
Scientific Research (KBN SPUB Nr 621/E-78/SPUB-M/CERN/P-03/DZ 298/2000
and KBN grant Nr 2 P03B 113 19). One of us (A.S.) acknowledges  
support from the Raymond and Beverly Sackler Visiting Chair in Exact Sciences
during his stay at Tel Aviv 
University.


\newpage

\newpage

\newpage

\newpage

\newpage

\newpage

\newpage

\noindent
\begin{thebibliography}{99}

\bibitem{low}
  F.E. Low, Phys. Rev. {\bf D\,12} (1975) 163.

\bibitem{bb} G. Bertsch {\it et al}, Phys. Rev. Lett. {\bf 47} (1981) 297.

\bibitem{bfgms} S.J. Brodsky, L.L. Frankfurt, J.F. Gunion, 
  A.H. Mueller, M.I. Strikman, Phys. Rev. {\bf D\,50} (1994) 3134.

\bibitem{fms}
  L.L. Frankfurt, G.A. Miller and M.I. Strikman, Phys. Lett.
  {\bf B304} (1993) 1;\\
  L.L. Frankfurt, G.A. Miller and M.I. Strikman,
  Ann. Rev. Nucl. Part. Sci. {\bf 45} (1994) 501.

\bibitem{bbfs}
  B. Bl\"{a}ttel, G. Baym, L.L. Frankfurt and M.I. Strikman,
  Phys. Rev. Lett. {\bf 70} (1993) 896.

\bibitem{fks}
  L.L. Frankfurt, W. K\"opf and  M.I. Strikman,
  Phys. Rev. {\bf D54} (1996) 3194.

\bibitem{frs}
  L.L. Frankfurt, A.V. Radyushkin and M.I. Strikman, Phys. Rev.
  {\bf D55} (1997) 98.
 
\bibitem{Sokolov} M.D. Sokoloff {\it et al}, Phys. Rev. Lett. {\bf 57}
(1986) 3003.

\bibitem{fgkss} L.L. Frankfurt, V. Guzey, W. K\"{o}pf, M. Sargsian
and M.I. Strikman, {\em Color Transparency and Color Opacity in 
Coherent Production of Vector Mesons off Light Nuclei at Small
$x$}, hep-ph/9608492.

\bibitem{rapgap} L.L. Frankfurt and M.I. Strikman, Phys. Lett.
  {\bf B382} (1996) 6.

\bibitem{almu}
A.H. Mueller, Proc. of the XVII Rencontre de Moriond, 1982, 
ed. J. Tr\^{a}n Thanh V\^{a}n,
Editions Fronti\`{e}res, Gif-sur-Yvette, France, 1982, p.13.

\bibitem{compass} G. Baum {\it et al}, 
{\it COMPASS, A Proposal 
for a Common Muon and
Proton Apparatus for Structure and Spectroscopy\/},\\ 
CERN/SPSLC/96-14, SPSC/P\,297 (1996) and CERN/SPSLC 96-30 (1996),\\
  http://wwwcompass.cern.ch/compass/proposal/welcome.html.

\bibitem{bradamante} F. Bradamante, Prog. Part. Nucl. Phys.
{\bf 44} (2000) 339.

\bibitem{RPW} T. Renk, G.Piller and W. Weise, {\em Coherence Effects in
Diffractive Electroproduction of $\rho ^0$ Mesons from Nuclei},
hep-ph/0008109, submitted to Nucl. Phys. A.

\bibitem{HERMES} HERMES Collab., K. Ackerstaff {\it et al}, 
Phys. Rev. Lett. {\bf 82} (1999) 3025.

\bibitem{collins} P.D.B. Collins, {\it  Introduction to Regge theory},
Cambridge Univ. Press, Cambridge, 1977.

\bibitem{Fra98} L.L. Frankfurt, M.V. Polyakov and M.I. Strikman, {\em
  N$\rightarrow \Delta$ DVCS, exclusive DIS processes and skewed quark
  distributions in large N$_c$ limit}, Workshop on Jefferson
  Lab. Physics and Instrumentation with 6-12 GeV Beams and Beyond, Newport
  News, Virginia, June 1998, hep-ph/9808449. 

\bibitem{h1} H1 Collab., C. Adloff {\it et al}, Eur. Phys. J. {\bf C13}
(2000) 371.

\bibitem{zeus} ZEUS Collab., J. Breitweg {\it et al}, Eur. Phys. J. {\bf C6}
(1999) 603.

\bibitem{Fra98b} L.L. Frankfurt and M.I. Strikman,
 Eur. Phys. J. {\bf A5} (1999) 293.

\bibitem{AGS} A.S. Carroll {\it et al}, Phys. Rev. Lett. {\bf 61} (1988)
1698.

\bibitem{SLAC} N.C.R. Makins {\it et al}, Phys. Rev. Lett. {\bf 72} 
(1994) 1986;\\
 T.G. O'Neil {\it et al}, Phys. Lett. {\bf B351} (1995) 87.

\bibitem{NMCJpsi} NMC, P. Amaudruz {\it et al}, Nucl. Phys. {\bf B371}
(1992) 553;\\
NMC, M. Arneodo {\it et al}, Phys. Lett. {\bf B332} (1994) 195.

\bibitem{E665} E665 Collab., M.R. Adams {\it et al}, 
Phys. Rev. Lett. {\bf 74} (1995) 1525. 

\bibitem{nmc} NMC, M. Arneodo {\it et al}, Nucl. Phys. {\bf B429}
  (1994) 503.

\bibitem{Wei97} E791 Collab., E.M. Aitala {\it et al}, 
Phys. Rev. Lett. {\bf 86} (2001) 4773;\\
E791 Collab., E.M. Aitala {\it et al},
Phys. Rev. Lett. {\bf 86} (2001) 4768.

\bibitem{memo2000} A. Sandacz, {\em Hard Exclusive Meson Production
at COMPASS}, COMPASS Note 2000-1, 
http://wwwcompass.cern.ch/compass/notes.

\bibitem{pmmpsv} J. Pochodzalla, 
L. Mankiewicz, M. Moinester, G. Piller, A. Sandacz, M. Vanderhaeghen, 
{\it Exclusive Meson Production at COMPASS}, hep-ph/9909534.

\bibitem{smcrho} A. Tripet, presented at
the 7th International Workshop on Deep Inelastic Scattering and QCD,
 Zeuthen, Germany, 19-23 April 1999; Proceedings, J. Bl\"{u}mlein
and T. Riemann, Nucl. Phys. B, Proc. Suppl. {\bf 79} (1999) 529.

\bibitem{nmc1} NMC, P. Amaudruz {\it et al}, Z. Phys. {\bf C54} (1992) 239.    

\bibitem{kurek} K. Kurek, {\it QED Radiative Corrections in
Exclusive $\rho ^0$ Leptoproduction}, DESY-96-209, hep-ph/9609240.

\bibitem{SW} K. Schilling and G. Wolf, Nucl. Phys. {\bf B61} (1973) 381.

\bibitem{schildknecht} D. Schildknecht, G.A. Schuler and B. Surrow,
Phys. Lett. {\bf B449} (1999) 328.

\bibitem{jackson} J.D. Jackson, Nuovo Cimento {\bf 34} (1964) 1644.

\bibitem{KM} K.S. K\"{o}lbig and B. Margolis, Nucl. Phys. {\bf B6} (1968) 85.

\bibitem{angdis} T.H. Bauer {\it et al}, Rev. Mod. Phys. {\bf 50} (1978)
261;\\
J. Ballam {\it et al}, Phys. Rev. {\bf D5} (1972) 545;\\
P. Joos {\it et al}, Nucl. Phys. {\bf B113} (1976) 53.

\bibitem{zeus1} ZEUS Collab., J. Breitweg {\it et al}, Eur. Phys. J. {\bf C12}
(2000) 393.

\bibitem{radii} H. de Vries {\it et al}, Atomic Data and Nucl. Data
Tables {\bf 36} (1987) 495.

\bibitem{adam} A. Mielech (COMPASS), private communication.

\bibitem{off} X. Ji, Phys. Rev. Lett. {\bf 78} (1997) 610;
Phys. Rev. {\bf D 55} (1997) 7114;\\
A.V. Radyushkin, Phys. Lett. {\bf B380} (1996) 417;
Phys. Rev. {\bf D 56} (1997) 5524;\\
J.C. Collins, L. Frankfurt and M. Strikman,
Phys. Rev. {\bf D 56} (1997) 2982.

\end{thebibliography}
\end{document}